# Sharp Load Thresholds for Cuckoo Hashing

Nikolaos Fountoulakis and Konstantinos Panagiotou*

Max-Planck-Institute for Informatics
Saarbrücken, Germany

**Abstract**

The paradigm of many choices has influenced significantly the design of efficient data structures and, most notably, hash tables. Cuckoo hashing is a technique that extends this concept. There, we are given a table with $n$ locations, and we assume that each location can hold one item. Each item to be inserted chooses randomly $k \geq 2$ locations and has to be placed in any one of them. How much load can cuckoo hashing handle before collisions prevent the successful assignment of the available items to the chosen locations? Practical evaluations of this method have shown that one can allocate a number of elements that is a large proportion of the size of the table, being very close to 1 even for small values of $k$ such as 4 or 5.

In this paper we show that there is a critical value for this proportion: with high probability, when the amount of available items is below this value, then these can be allocated successfully, but when it exceeds this value, the allocation becomes impossible. We give explicitly for each $k \geq 2$ this critical value. This answers an open question posed by Mitzenmacher (*ESA '09*) and underpins theoretically the experimental results. Our proofs are based on the translation of the question into a hypergraph setting, and the study of the related typical properties of random $k$-uniform hypergraphs.

## 1 Introduction

A fundamental data structure that is omnipresent in various computer science applications is the dictionary. A dictionary typically supports the following three basic operations: insertion and deletion of elements as well as membership queries. One way to design dictionaries is by using hash-based techniques, which become highly efficient when the power of multiple choices is exploited. In particular, if each item has the choice among several different locations in the table, and simultaneously the items may be moved dynamically to different locations, very efficient data structures that achieve small lookup times and high memory utilizations can be designed. In this work we will focus on a variation of this concept that is called *Cuckoo Hashing*.

Cuckoo hashing was introduced by Pagh and Rodler in [23]. The general setting considered here is a slight variation of it, as defined by Fotakis, Pagh, Sanders and Spirakis in [15]. We are given a table with $n$ locations, and we assume that each location can hold only one item. Further generalizations where two or more items can be stored have also been studied, see e.g. [11, 5, 14], but we will not treat those cases. Moreover, there is a set $\mathcal{I}$ of items, such that each $x \in \mathcal{I}$ is assigned $k \geq 2$ random distinct locations $\ell_1(x), \ldots, \ell_k(x)$. This assumption may be somehow idealized, as exponentially many bits would be needed to store such fully random sets of locations. However, there is theoretical evidence that even "simple" hash functions can be sufficient in practice, provided that the underlying data stream fulfills certain natural conditions; we refer the reader to the papers [21] by Mitzenmacher and Vadhan and [10] by Dietzfelbinger and Schellbach, and the references therein.

---

*This author was supported by the Humboldt Foundation.

A natural question in cuckoo hashing is the following. As the number of available items increases, it becomes more and more unlikely that all of them can be inserted into the table such that each item is assigned to one of its $k$ desired locations. In other words, if $|\mathcal{I}|$ is "small", then with high probability, i.e., with probability arbitrarily close to 1 when $n$ becomes large, there is such an assignment to the locations in the table that respects the $k$ choices of the items. On the other hand, if $|\mathcal{I}|$ becomes "large", then such an assignment does not exist with high probability (trivially, this happens at the latest when $n+1$ items are available). The important question is whether there is a critical size for $\mathcal{I}$ where the probability for the existence of a valid assignment drops instantaneously in the limiting case from 1 to 0, i.e., whether there is a *load threshold* for cuckoo hashing. This question is at the core of this article. More precisely, we shall say that a value $c_k^*$ is the load threshold for cuckoo hashing with $k$ choices for each element if

$$\mathbb{P}\left(\begin{array}{c}\text{there is an assignment of } |\mathcal{I}| = \lfloor cn \rfloor \text{ items to a table}\\ \text{with } n \text{ locations that respects the choices of all items}\end{array}\right) \stackrel{n \to \infty}{\Rightarrow} \begin{cases} 1, & \text{if } c < c_k^*, \\ 0, & \text{if } c > c_k^* \end{cases}. \quad (1.1)$$

It is not *a priori* clear whether such a critical value $c_k^*$ exists at all. However, as a warm-up example let us consider the case $k = 2$, that is, each item has two preferred random locations. Then there is a natural analogy to random graphs. Indeed, let us think of the $n$ locations as the vertices of the graph, and of the items as edges, which encode the two choices for each item. What we obtain in this way is the Erdős-Rényi random (multi-)graph $G_{n,m}$, where $m = |\mathcal{I}|$. Moreover, it easy to see by applying Hall's Theorem that $G_{n,m}$ has no subgraph with more edges than vertices if and only if the corresponding items can be assigned to the corresponding locations such that the choices of all items are respected. It is well-known that the property "$G_{n,m}$ has a subgraph with more edges than vertices" coincides with the appearance of a *giant connected component* that contains a linear fraction of the vertices, see e.g. the detailed book [17]. As the latter is known to happen around $m = n/2$, we readily obtain that the load threshold for cuckoo hashing and $k = 2$ is at $c_2^* = 1/2$. Particularly, roughly *half* of the table can be filled, while respecting the choices of all items.

In the case $k \geq 3$ things get significantly more complex, and there are almost no precise results about the existence and the value of the corresponding load threshold $c_k^*$ (see the next section for a detailed account on the related work and previously known bounds). However, experiments in [15] suggest that e.g. $c_3^* \approx 0.91$, $c_4^* \approx 0.97$ and $c_5^* \approx 0.99$. In other words, three choices allow already an utilization of 91% of the table, and e.g. five choices are sufficient to fill the table almost completely. The main result of this work is the following theorem, which says that for all $k \geq 3$ the load threshold $c_k^*$ exists, and determines its value. This in particular answers a question by Mitzenmacher [20] related to the maximum load that can be achieved when $k \geq 3$ choices are allowed for the elements.

**Theorem 1.1.** *For any integer $k \geq 3$ let $\xi^*$ be the unique solution of the equation*

$$k = \frac{\xi^*(e^{\xi^*} - 1)}{e^{\xi^*} - 1 - \xi^*}. \quad (1.2)$$

Then $c_k^* = \frac{\xi^*}{k(1-e^{-\xi^*})^{k-1}}$ is the load threshold for cuckoo hashing with $k$ choices per element. In particular, if there are $\lfloor cn \rfloor$ items, then the following holds with probability $1 - o(1)$.

1. If $c < c_k^*$, then there is an assignment of the items to a table with $n$ locations that respects the choices of all items.

2. If $c > c_k^*$, then such an assignment does not exist.

Numerically we obtain for example that $c_3^* \doteq 0.917, c_4^* \doteq 0.976$ and $c_5^* \doteq 0.992$, where "$\doteq$" indicates that the values are truncated to the last digit shown. Note that this matches in all cases the empirically observed values from [15] up to the second decimal digit. Moreover, a simple calculation reveals that $c_k^* = 1 - e^{-k} + o(e^{-k})$ for $k \to \infty$, i.e., the utilization rate of the table approaches one exponentially fast in $k$.

In the remainder of this section we first briefly survey some known results about cuckoo hashing in various settings, and we refer the reader to the survey [20] by Mitzenmacher, which contains a much more complete treatment. Then we give a high-level overview of our proofs, and point out the novelties in our approach and the differences to past approaches.

**Related Results** A breakthrough in hashing with multiple choices was achieved by Azar, Broder, Karlin, and Upfal [2], who studied the following balls-into-bins game. Suppose that we sequentially place $n$ balls into $n$ bins such that for each ball we choose randomly two bins, and place the ball into the least full. Then this yields with high probability an *exponential* improvement regarding the maximum number of balls in any bin, compared to the game where each ball is just placed into a random bin. Particularly, in [2] it was shown that the maximum load grows like $\frac{\log \log n}{\log 2}$, which in most practical scenarios is very small. A variant of this scheme was later considered by Vöcking [24], who showed even smaller upper bounds for the resulting maximum load.

Cuckoo hashing is a further development of the above idea [23]. There, the items not only have multiple choices, but are also allowed to change their current location on demand. This might of course lead to a sequence of (recursive) movements of the items, until the newly arrived and all preceding items are inserted successfully into the table, if this was possible.

The case of $k = 2$ choices is theoretically well-understood. As already mentioned, there is a natural correspondence to random graphs, which makes the analysis tractable; we refer the reader to [8] and [12] for further details. However, the cases $k \geq 3$ are not that well-understood.

Generally we can distinguish between two different settings of cuckoo hashing: the *online* case, where items are inserted or deleted over time from the data structure, and the *offline* case, where all required items are available already in the beginning. This paper is devoted to the study of the offline setting, and the important question in this context is about the maximum number of initially available items that can be inserted in the table with high probability, so that the choices of all items are respected (this is precisely the load threshold as defined above in (1.1)). The work [3] by Batu, Berenbrink and Cooper considers, among other things, the case $k = 4$, and derives an explicit expression for $c_4^*$, which matches the value in Theorem 1.1. Moreover, they derive upper bounds for the cases $k = 3$ and $k = 5$. Moreover, (non-tight) lower bounds on $c_k^*$ for large $k$ have been achieved with the use of matrix techniques, see the paper [9] by Dietzfelbinger and Pagh.

In the online setting many different approaches have been explored, see e.g. [11, 15, 16]. Particularly, the popular "random-walk" insertion algorithm has attracted much attention, and Frieze, Melsted and Mitzenmacher [16] showed at most polylogarithmic insertion time for sufficiently large $k$ and appropriate (but not optimal) maximum loads of the table. We refer the reader to [20] for a more detailed treatment.

Finally, a different line of research in cuckoo hashing aims at a *deamortization* of the time needed for the insertion of the elements. Note that while a membership query can be answered by checking at most $k$ locations in the table, an insertion procedure (like e.g. the random-walk algorithm) could take significantly longer time. The papers [1] and [19] contain a very detailed treatment of this topic, and we refer the reader to them.

**Proof Methods** Typically, but not exclusively, cuckoo hashing problems are modeled with a bipartite graph: the parts represent the locations and the items, respectively, and the edges represent the choices of the items. It is easy to see that the existence of a matching in this graph that covers all item-nodes is equivalent to the existence of an assignment of all items to the locations such that the choices are respected. In other words, we are seeking the threshold for the existence of such a matching.

Our approach is also based on random graphs, but we adopt a different viewpoint. Particularly, we model cuckoo hashing with random (uniform) *hyper*graphs, similarly to the case $k = 2$. More precisely, the $n$ vertices represent the locations, and each available item corresponds to a hyperedge that contains $k$ vertices. It is not very difficult to see (see also Section 2) that the existence of an assignment of all items to the locations such that the choices are respected is guaranteed to exist, if the underlying hypergraph does not contain any subgraph with more edges than vertices. Otherwise, an assignment is not possible.

Unfortunately, unlike in the case $k = 2$, studying the emergence of a giant component is not sufficient. In fact, it turns out that the giant component, shortly after it appears in a random $k$-uniform hypergraph, has significantly less edges than vertices. What we are interested in here is the threshold for the appearance of a dense subgraph, i.e., a subgraph that contains at least as many edges as vertices. An upper bound for this threshold can be obtained by considering the so-called *core*, which is the maximum subgraph that has minimum degree at least two. By using well-known results by Cooper [6] and Kim [18] we infer in Theorem 2.2 that the sought threshold is at most $c_k^*$. The proof can be found in Section 3.1.

The main contribution of our work is the corresponding lower bound, i.e., to show that whenever $c < c_k^*$ there are no dense subgraphs in the random $k$-uniform hypergraph with $n$ vertices and $\lfloor cn \rfloor$ edges. This was shown already by Bohman and Kim [4] for the case $k = 4$. However, most of their proof is tailored specifically to this case, and does not extend to the whole range $k \geq 3$. Our proof is heavily based on the use of the probabilistic method and tools from large deviation theory (see the book [13] by Ellis and [7] by Dembo and Zeitouni), and boils down to a delicate maximization of a function that involves entropic terms as well as terms that arise from various probability estimates that we perform. We refer the reader to Section 4.1 for further details.

## 2 Proof of the Main Result

As already mentioned, we model cuckoo hashing using random hypergraphs: the $n$ vertices represent the locations, and each available item is a hyperedge that contains $k$ vertices. Let us become a little more precise. Let $H_{n,m,k}^*$ denote a hypergraph that is drawn uniformly at random from the set of multigraphs with $n$ vertices and $m$ edges, where each edge contains $k$ vertices. Then, an instance of $H^*_{n,m,k}$ corresponds precisely to a cuckoo hashing scenario, where the table consists of $n$ locations, there are $m$ items in total, and each items chooses $k$ random locations as specified above.

Note that $H^*_{n,m,k}$ may have multiple edges, which happens when two items choose the same $k$ locations. It will be more convenient to work on a slightly different random graph model, which we denote by $H_{n,m,k}$. There, multiple edges are forbidden, i.e., $H_{n,m,k}$ is a hypergraph drawn uniformly at random from the set of all simple hypergraphs with $n$ vertices and $m$ edges. The next proposition says that the two models are essentially equivalent.

**Proposition 2.1.** *Let $\mathcal{P}$ be a property of simple hypergraphs. Then, for any $c > 0$ and $k \geq 3$*

$$\mathbb{P}\left(H_{n,\lfloor cn \rfloor,k} \in \mathcal{P}\right) = \mathbb{P}\left(H^*_{n,\lfloor cn \rfloor,k} \in \mathcal{P}\right)(1 + o(1)).$$

*Proof.* The expected number of pairs of edges that are incident to the same set of vertices in $H^*_{n,\lfloor cn \rfloor,k}$ is at most $(cn)^2 \cdot \binom{n}{k} \cdot \binom{n}{k}^{-2} = o(1)$. This implies that $H^*_{n,\lfloor cn \rfloor,k}$ is simple with probability $1 - o(1)$, and the statement follows immediately. $\square$

In what follows we will be referring to a hyperedge of size $k$ as a *k-edge* and we will be calling a hypergraph where all of its hyperedges are of size $k$ a *k-graph*.

Suppose now that we are given an empty table with $n$ locations, and $m$ items together with $k$ choices for each one of them. Note that an assignment of the items to the locations of the table such that every item gets assigned to one of its preferred locations is possible, if for the corresponding $k$-graph $H$ the following condition is satisfied: every one of the induced subgraphs of $H$ has less $k$-edges than vertices. Indeed, every one of the edges corresponds to an item that has been allocated to precisely one location in the table, or equivalently, to one vertex of the subgraph. But the existence of more $k$-edges than vertices within an induced subgraph implies that the number of items that have been allocated to a certain set of table locations exceeds the number of admissible locations – clearly this is impossible. We call the ratio of the number of edges of a $k$-graph over its number of vertices the *density* of this $k$-graph.

It turns out that this necessary condition is also sufficient – see the proof of Theorem 1.1 below. In other words, the crucial parameter that determines whether an assignment of the items to the locations of the table is possible is the maximal density of an induced subgraph. The next theorem says that if the number of items exceeds $c_k^* n$, then there is a subgraph with density $> 1$. Before we state it, let us introduce some additional notation. We define the *core* of a hypergraph $H$ to be the maximum subgraph of $H$ that has minimum degree at least 2; if such a subgraph does not exist then we say that the core *is empty*. The core of random hypergraphs and its structural characteristics have been studied quite extensively in recent years – see for example the papers by Cooper [6] or Molloy [22].

**Theorem 2.2.** *Let $c_k^*$ be defined as in Theorem 1.1. If $c > c_k^*$, then with probability $1 - o(1)$ the core of $H_{n,cn,k}$ has density greater than 1.*

This theorem is not very difficult to prove, given the results in [6] and [18]. However, it requires some technical work, which is accomplished in Section 3. The heart of this paper is devoted to the "subcritical" case, where we show that the above result is essentially tight.

**Theorem 2.3.** *Let $c_k^*$ be defined as in Theorem 1.1. If $c < c_k^*$, then with probability $1 - o(1)$ all subgraphs of $H_{n,cn,k}$ have density smaller than 1.*

The proof of this theorem is technically more challenging and it is spread over the remaining sections. With Theorems 2.2 and 2.3 at hand we are in a position to give the proof of our main result about cuckoo hashing.

*Proof of Theorem 1.1.* Let us construct an auxiliary bipartite graph $B = (I, L; E)$, where $I$ represents the $m$ items, $L$ represents the $n$ locations, and $\{i, \ell\} \in E$ if $\ell$ is one of the $k$ preferred locations for item $i$. Note that it is possible to assign all items to locations such that each item is assigned to one of its preferred locations if and only if there is a matching in $B$ that covers all vertices in $I$. By Hall's Theorem such a matching exists if and only if for all $I' \subseteq I$ we have that $|I'| \leq |\Gamma(I')|$, where $\Gamma(I')$ denotes the set of neighbors of the vertices in $I'$ inside $L$.

As a next step, let us describe more precisely the quantity $\Gamma(I')$. If we think of cuckoo hashing in terms of the corresponding random hypergraph $H_{n,m,k}$, then $\Gamma(I')$ is precisely the set of vertices that are contained in the hyperedges that correspond to the items in $I'$. So, if $c < c_k^*$, Theorem 2.3 guarantees that with high probability for all $I'$ we have $|\Gamma(I')| > |I'|$, and therefore a matching exists. On the other hand, if $c > c_k^*$, then there is with high probability an $I'$ such that $|\Gamma(I')| < |I'|$; choose for example $I'$ to be the set of items that correspond to the edges in the core of $H_{n,m,k}$. Hence a matching does not exist in this case, and the proof is completed. □

## 3 Properties of random $k$-graphs

The aim of this section is to determine the value $c_k^*$ and to prove Theorem 2.2. Moreover, we will introduce some known facts and tools that will turn out to be very useful in the study of random hypergraphs, and will be used later on in the proof of Theorem 2.3 as well.

### (More) Models of random $k$-graphs

For the sake of convenience we will carry out our calculations in the $H_{n,p,k}$ model of random $k$-graphs. This is the "higher-dimensional" analogue of the well-studied $G_{n,p}$ model, where each possible (2-)edge is included independently with probability $p$. More precisely, given $n \geq k$ vertices we obtain $H_{n,p,k}$ by including each $k$-tuple of vertices with probability $p$, independently of every other $k$-tuple.

Standard arguments show that if we adjust $p$ suitably, then the $H_{n,p,k}$ model is essentially equivalent to the $H_{n,cn,k}$ model. Roughly speaking, if we set $p = ck/\binom{n-1}{k-1}$, then $H_{n,p,k}$ is expected to have $p\binom{n}{k} = cn$ edges. In fact, much more is true. Let $\mathcal{P}$ be a *convex* hypergraph property, that is, whenever we have three hypergraphs on the same vertex set $H_1, H_2, H_3$ such that $H_1 \subseteq H_2 \subseteq H_3$ and $H_1, H_3 \in \mathcal{P}$, then $H_2 \in \mathcal{P}$ as well. (We also assume that $\mathcal{P}$ is closed under automorphisms.) Clearly any monotone property is also convex and, therefore, in particular the properties examined in Theorem 2.3. The following proposition is a generalization of Proposition 1.15 from [17, p.16] and its proof is very similar to the proof of that – so we omit it.

**Proposition 3.1.** *Let $\mathcal{P}$ be a convex property of hypergraphs, and let $p = ck/\binom{n-1}{k-1}$, where $c > 0$. If $\mathbb{P}(H_{n,p,k} \in \mathcal{P}) \to 1$ as $n \to \infty$, then $\mathbb{P}(H_{n,cn,k} \in \mathcal{P}) \to 1$ as well.*

**Working on the core of $H_{n,p,k}$ – the Poisson cloning model**

Recall that the core of a hypergraph is its maximum subgraph that has minimum degree at least 2. At this point we introduce the main tool for our analysis of the core of $H_{n,p,k}$. This is the so-called *Poisson cloning model* that was introduced by Kim [18] and was used for a variety of problems. Our treatment here was inspired by the analysis of Bohman and Kim [4] in the context of Achlioptas processes.

The Poisson cloning model $\widetilde{H}_{n,p,k}$ for $k$-graphs with $n$ vertices and parameter $p \in [0,1]$ is defined as follows. Consider a set of $n$ vertices $V_n$ and consider also a family, indexed by this set, of i.i.d. Poisson random variables with parameter $\lambda := p\binom{n-1}{k-1}$. For each $v \in V_n$ let $d(v)$ denote the corresponding random variable from this family. Then $\widetilde{H}_{n,p,k}$ is constructed in three steps as follows. First, for every $v \in V_n$ the degree of $v$ is a random variable and equals $d(v)$. Second, for each such $v$ we generate $d(v)$ copies, which we call $v$-*clones* or simply *clones*, and choose uniformly at random a matching from all perfect $k$-matchings on the set of all clones. Note that such a matching may not exist – in this case we choose a random matching that leaves less than $k$ clones unmatched. Finally, we generate $\widetilde{H}_{n,p,k}$ by contracting the clones to vertices, i.e., by projecting the clones of $v$ to $v$ itself for every $v \in V_n$.

Note that the last two steps in the above procedure are together equivalent to the *configuration model* $H_{\mathbf{d},k}$, where $\mathbf{d} = (d_1, \ldots, d_n)$, for random hypergraphs with degree sequence $\mathbf{d}$. In other words, $H_{\mathbf{d},k}$ is a random multigraph where the $i$th vertex has degree $d_i$.

The following statement is implied by [18, Theorem 1.1], and says that the study of $H_{n,p,k}$ may be reduced to the study of the Poisson cloning model.

**Theorem 3.2.** *If $\mathbb{P}\left(\widetilde{H}_{n,p,k} \in \mathcal{P}\right) \to 0$ as $n \to \infty$, then $\mathbb{P}\left(H_{n,p,k} \in \mathcal{P}\right) \to 0$ as well.*

The next result that we shall exploit gives a precise description of the core of $\widetilde{H}_{n,p,k}$. Particularly, Theorem 6.2 in [18] implies the following statement, where we write "$x \pm y$" for the interval of numbers $(x-y, x+y)$.

**Theorem 3.3.** *Let $\lambda_2 := \min_{x>0} \frac{x}{(1-e^{-x})^{k-1}}$. Moreover, let $\bar{x}$ be the largest solution of the equation $x = (1-e^{-xck})^{k-1}$, and set $\xi := \bar{x}ck$. Assume that $ck = p\binom{n-1}{k-1} = \lambda_2 + \sigma$, where $\sigma > 0$ is fixed. Then, for any $0 < \delta < 1$ the following is true with probability $1 - o(1)$. If $\tilde{n}_2$ denotes the number of vertices in the core of $\widetilde{H}_{n,p,k}$, then*

$$\tilde{n}_2 = (1 - e^{-\xi} - \xi e^{-\xi})n \pm \delta n.$$

*Furthermore, the core itself is distributed like the Poisson cloning model on $\tilde{n}_2$ vertices, where the Poisson random variables are* conditioned on being at least two *and have parameter $\Lambda_{c,k}$, where $\Lambda_{c,k} = \xi + \beta$, for some $|\beta| \le \delta$.*

In what follows, we say that a random variable is a 2-truncated Poisson variable, if it is distributed like a Poisson variable, conditioned on being at least two. We immediately obtain the following corollary.

**Corollary 3.4.** *Let $n_2$ and $m_2$ denote the number of vertices and edges in the core of $H_{n,p,k}$. Then, for any $0 < \delta < 1$, with probability $1 - o(1)$,*

$$n_2 = (1 - e^{-\xi} - \xi e^{-\xi})n \pm \delta n \quad \text{and} \quad m_2 = \frac{1}{k}\xi(1 - e^{-\xi})n \pm \delta n,$$

*where $\xi = \bar{x}ck$ and $\bar{x}$ is the largest solution of the equation $x = (1 - e^{-xck})^{k-1}$. The same is true for the quantities $\tilde{n}_2$ and $\tilde{m}_2$ of $\widetilde{H}_{n,p,k}$.*

*Proof.* The statement about $n_2$ follows immediately from the first part of Theorem 3.3 together with Theorem 3.2. The second part is not substantially more difficult. We will use the statement about the distribution of the core of $\widetilde{H}_{n,p,k}$ in Theorem 3.3, that is, the 2-truncated Poisson cloning model, and we transfer this result to $H_{n,p,k}$ using again Theorem 3.2.

Let us condition on certain values of $\tilde{n}_2$ and $\Lambda_{c,k}$ that lie in the intervals stated in Theorem 3.3. Then the total degree of the core of $\widetilde{H}_{n,p,k}$ is the sum of independent 2-truncated Poisson random variables $d_1, \ldots, d_{\tilde{n}_2}$ with parameters in $\Lambda_{c,k} \in \xi \pm \delta$. Set $D := \sum_{i=1}^{\tilde{n}_2} d_i$. A simple calculation shows that $\text{Var}(D) = \Theta(\tilde{n}_2)$. Therefore, Chebyschev's inequality yields

$$\mathbb{P}\left(|D - \mathbb{E}(D)| > \tilde{n}_2^{2/3}\right) = O\left(\tilde{n}_2^{-1/3}\right).$$

Also, $\mathbb{E}(D) = (1 \pm c\delta) \cdot \frac{\xi(1-e^{-\xi})}{1-e^{-\xi}-\xi e^{-\xi}} \cdot \tilde{n}_2$, for some appropriate $c > 0$. Therefore, by averaging over all choices of $\tilde{n}_2$ in the interval stated in Theorem 3.3, we obtain that $\tilde{m}_2 = \frac{1}{k}\xi(1-e^{-\xi})n \pm c'\delta n$, for some $c' > 0$. The proof completes by choosing the initial $\delta$ as $\delta/c'$, and by applying Theorem 3.2. $\square$

## 3.1 Proof of Theorem 2.2 and the Value of $c_k^*$

In this section we will prove Theorem 2.2, i.e., we will show that the core of $H_{n,p,k}$ has density at least one if $p = ck/\binom{n-1}{k-1}$ and $c > c_k^*$. Let $0 < \delta < 1$, and denote by $n_2$ and $m_2$ the number of vertices and edges in the core of $H_{n,p,k}$. By applying Corollary 3.4 we obtain that with probability $1 - o(1)$

$$n_2 = (1 - e^{-\xi} - \xi e^{-\xi})n \pm \delta n \quad \text{and} \quad m_2 = \frac{1}{k}\xi(1-e^{-\xi})n \pm \delta n,$$

where $\xi = \bar{x}ck$ and $\bar{x}$ is the largest solution of the equation $x = (1 - e^{-xck})^{k-1}$. The value of $c_k^*$ is then obtained by taking $n_2 = m_2$, and ignoring the additive error terms. The above values imply that the critical $\xi^*$ is given by the equation

$$1 - e^{-\xi^*} - \xi^* e^{-\xi^*} = \frac{\xi^*}{k}(1 - e^{-\xi^*}) \implies k = \frac{\xi^*(1-e^{-\xi^*})}{1-e^{-\xi^*}-\xi^* e^{-\xi^*}}.$$

This is precisely (1.2). So, $k$ determines $\xi^*$ and $\bar{x}$ satisfies $\bar{x} = (1-e^{-\bar{x}ck})^{k-1} = (1-e^{-\xi^*})^{k-1}$. Therefore, the critical density is

$$c_k^* := \frac{1}{k}\frac{\xi^*}{(1-e^{-\xi^*})^{k-1}}. \tag{3.1}$$

*Proof of Theorem 2.2.* The above calculations imply that with probability $1 - o(1)$, for any $0 < \delta < 1$

$$\frac{m_2}{n_2} = \frac{1}{k}\frac{\xi(1-e^{-\xi})}{1-e^{-\xi}-\xi e^{-\xi}} \pm 2\delta.$$

Moreover, if $c = c_k^*$, then $m_2/n_2 = 1 \pm 2\delta$. To complete the proof it is therefore sufficient to show that $m_2/n_2$ is an increasing function of $c$. Note that the ratio $\frac{\xi(1-e^{-\xi})}{1-e^{-\xi}-\xi e^{-\xi}}$ is the expected value of a 2-truncated Poisson random variable with parameter $\xi$, which is known (and easily seen) to be increasing in $\xi$. Recall that $\xi = \bar{x}ck$. We conclude the proof by showing the following claim.

**Claim 3.5.** *The quantity $\xi = \bar{x}ck$ is increasing with respect to $c$. So, with probability $1 - o(1)$*

$$\frac{m_2}{n_2} < 1 \ , \ \text{if } c < c_k^* \qquad \text{and} \qquad \frac{m_2}{n_2} > 1 \ , \ \text{if } c > c_k^*.$$

Indeed, recall that $\bar{x}$ satisfies $\bar{x} = (1 - e^{\bar{x}ck})^{k-1}$. Equivalently, $\bar{x}ck = ck(1 - e^{\bar{x}ck})^{k-1}$. We have

$$ck = \frac{\xi}{(1 - e^{-\xi})^{k-1}}. \tag{3.2}$$

An easy calculation shows that the derivative of the function on the right-hand side has no positive roots:

$$\left[\frac{\xi}{(1-e^{-\xi})^{k-1}}\right]' = \frac{(1-e^{-\xi})^{k-2}\left(1 - e^{-\xi} + e^{-\xi}\xi(k-1)\right)}{(1-e^{-\xi})^{2(k-1)}} \stackrel{(\xi>0)}{>} 0,$$

That is, the function $\frac{\xi}{(1-e^{-\xi})^{k-1}}$ is strictly increasing for $\xi > 0$ and, therefore, when $ck$ increases, then the root of (3.2), that is, the product $\bar{x}ck$, increases as well. □

## 4 Proof of Theorem 2.3

Let us begin with introducing some notation. For a hypergraph $H$ we will denote by $V_H$ its vertex set and by $E_H$ its set of edges. Additionally, $v_H$ and $e_H$ shall denote the number of elements in the corresponding sets. For $U \subset V_H$ we denote by $v_U$, $e_U$ the number of vertices in $U$ and the number of edges joining vertices only in $U$. Finally, $d_U$ is the total degree in $U$, i.e., the sum of the degrees in $H$ of all vertices in $U$.

We say that a subset $U$ of the vertex set of a hypergraph is 1-dense, if it has density at least one, i.e., the subgraph induced by $U$ has at least as many edges as vertices. Working towards the proof of Theorem 2.3, we begin with showing that whenever $c < 1$, $H_{n,cn,k}$ does not contain large 1-dense subsets, for any $k \geq 3$. In particular, we will first argue about sets with no more than $0.7n$ vertices; for them it is sufficient to use a first moment argument that is based on rough counting. For larger sets of vertices we need more sophisticated arguments regarding the structure of the core of $H_{n,cn,k}$ – we present those in the next subsection.

The following statement deals with the case $k \geq 5$; is not best possible, but it suffices for our purposes.

**Lemma 4.1.** *Let $k \geq 5, c < 1$. Then $H_{n,cn,k}$ contains no 1-dense subset with less than $0.7n$ vertices with probability $1 - o(1)$.*

*Proof.* The probability that an edge of $H_{n,cn,k}$ is contained completely in a subset $U$ of the vertex set is $\binom{|U|}{k}/\binom{n}{k} \leq (\frac{|U|}{n})^k$. Let $\frac{k}{n} \leq u \leq 0.7n$. Then

$$\mathbb{P}\left(\exists \text{ 1-dense subset with } un \text{ vertices}\right) \leq \binom{n}{un} \cdot \binom{cn}{un} u^{kun} \leq e^{n(2H(u) + ku \ln u)},$$

where $H(x) = -x \ln x - (1-x) \ln x$ denotes the entropy function. Note that the second derivative of the exponent in the expression above is $\frac{k-2+kx}{x(1-x)}$, which is positive for $x \in (0,1)$. Hence the exponent is convex, implying that it is maximized either at $u = k/n$ or at $u = 0.7$. Note that

$$2H(0.7) + k0.7\ln(0.7) \leq 2H(0.7) + 5 \cdot 0.7\ln(0.7) \leq -0.02$$

and that
$$2H\left(\frac{k}{n}\right) + \frac{k^2}{n}\ln\left(\frac{k}{n}\right) = -\frac{(k^2 - 2k)\ln n}{n} + O\left(\frac{1}{n}\right).$$
So, the minimum is obtained at $u = k/n$, and we conclude the proof with
$$\mathbb{P}\left(\exists \text{ 1-dense subset with } \leq 0.7n \text{ vertices}\right) \leq \sum_{k/n \leq u \leq 0.7} O(n^{-k^2+2k}) \leq O(n^{-14}).$$

□

We shall need the following stronger claim for random 3-uniform hypergraphs. The assumptions might look somehow artificial at this point, but will become very handy later on, cf. Proposition 4.3.

**Lemma 4.2.** *Let $H$ be a 3-graph, and call a set $U \subset V_H$ bad if*
$$e_U = |U| \quad \text{and} \quad \forall e \in E_H : |e \cap U| \neq 2.$$
*Then, for any $c \leq 0.95$*
$$\mathbb{P}\left(H_{n,cn,3} \text{ contains a bad subset } U \text{ with } \leq n/2 \text{ vertices}\right) = o(1).$$

*Proof.* Let $p = c'/\binom{n-1}{2}$, where $c' = 3 \cdot 0.95 \leq 2.85$. A simple application of Stirling's formula reveals
$$\mathbb{P}\left(H_{n,p,3} \text{ has exactly } cn \text{ edges}\right) = (1 + o(1))(2\pi cn)^{-1/2}.$$
As the distribution of $H_{n,cn,3}$ is the same as the distribution of $H_{n,p,3}$ conditioned on the number of edges being precisely $cn$ we infer that
$$\mathbb{P}\left(H_{n,cn,3} \text{ contains a bad subset } U \text{ with } \leq n/2 \text{ vertices}\right)$$
$$= O(n^{1/2}) \cdot \mathbb{P}\left(H_{n,p,3} \text{ contains a bad subset } U \text{ with } \leq n/2 \text{ vertices}\right).$$

To complete the proof it is therefore sufficient to show that the latter probability is $o(n^{-1/2})$. We accomplish this in two steps. Note that if a subset $U$ is bad, then certainly $|U| \geq 4$. Let us begin with the case $s := |U| \leq n^{1/4}$. There are at most $n^s$ ways to choose such a $U$, and at most $s^{3s}$ ways to choose the edges that are contained in $U$. Hence, the probability that $H_{n,p,3}$ contains a bad subset with at most $n^{1/4}$ vertices is bounded for large $n$ from above by
$$\sum_{s=4}^{n^{1/4}} n^s s^{3s} p^s = \sum_{s=4}^{n^{1/4}} \left(ns^3 \frac{c'}{\binom{n-1}{2}}\right)^s \leq \sum_{s=4}^{n^{1/4}} \left(n^{(1+3/4)-2} \cdot O(1)\right)^s \leq \sum_{s=4}^{n^{1/4}} \left(n^{-1/4+o(1)}\right)^s = n^{-1+o(1)}.$$

The case $n^{1/4} \leq |U| \leq n/2$ requires a more careful analysis. Let $n^{-3/4} \leq u \leq 1/2$, and denote by $H(x) = -x \ln x - (1-x)\ln(1-x)$ the entropy function. For future reference let us note that
$$\ln p = \ln\left(\frac{c'}{\binom{n-1}{2}}\right) = \ln\left(\frac{2c'}{n^2}\right) + \Theta(1/n). \tag{4.1}$$

By using the above notation, there are $\binom{n}{un} \leq e^{nH(u)}$ ways to select a $U$ with $un$ vertices. Moreover, the number of ways to choose the $un$ edges that are completely contained in $U$ is
$$\binom{\binom{un}{3}}{un} \leq \left(\frac{e\binom{un}{3}}{un}\right)^{un} \leq \left(\frac{e(un)^3}{6\,un}\right)^{un} = \exp\left\{un \ln\left(\frac{e(un)^2}{6}\right)\right\}.$$

Finally, the number of all possible edges with two endpoints in $U$ and one endpoint in $V_H \setminus U$ is $\binom{un}{2}(1-u)n$. By combining all these facts it readily follows that the probability $P_u$ that $H_{n,p,3}$ contains a bad subset $U$ with $un$ vertices is at most

$$P_u \leq \binom{n}{un}\binom{\binom{un}{3}}{un} p^{un}(1-p)^{\binom{un}{3}-un} \cdot (1-p)^{\binom{un}{2}(1-u)n}$$

$$\leq \exp\left\{n\left(H(u) + u\ln\left(\frac{e(un)^2}{6}\right) + u\ln p\right) - p\left(\binom{un}{3} - un + \binom{un}{2}(1-u)n\right)\right\}.$$

Using (4.1) and the definition of $p$ we obtain that uniformly for $n^{-3/4} \leq u \leq 1/2$

$$P_u \leq \exp\left\{n\left(H(u) + u\ln\left(\frac{ec'u^2}{3}\right) - c'\frac{u^3}{3} - c'u^2(1-u)\right) + \Theta(1)\right\}.$$

The derivative of the exponent with respect to $c'$ is given by $\frac{u(3+2c'u^2-3uc)}{3c'}$, which is easily seen to be positive whenever $c' < 3$ and $u \leq 1/2$. We infer that $P_u$ is maximized if we choose $c' = 2.85$. Moreover, an easy calculation reveals that the derivative of the exponent with respect to $u$ equals $\ln(c'u(1-u)) + 3 - \ln(3) - 2c'u(1-u)$. As the function $\ln(c'x(1-x))$ is monotone increasing for $0 \leq x \leq 1/2$ and the function $-2c'x(1-x)$ is monotone decreasing, there is at most one $n^{-3/4} \leq u_0 \leq 1/2$ where the derivative of the exponent vanishes. Moreover, it is easily seen that for $c = 2.85$ the derivative of the exponent is negative at $u = n^{-3/4}$ and positive at $u = 1/2$. Hence $u_0$ is a global minimum, and $P_u$ is maximized either at $u = n^{-3/4}$ or at $u = 1/2$. Elementary algebra then yields that the left point is the right choice, giving the estimate $P_u = o(2^{-n^{1/4}})$, and the proof concludes by adding up this expression for all admissible $n^{-3/4} \leq u \leq 1/2$. $\square$

## 4.1 Subgraphs of the 2-Core

Our general proof strategy for Theorem 2.3 is as follows. Suppose that $H_{n,cn,k}$ contains a 1-dense subset of vertices, and let $U$ be such a minimal (with respect to the number of vertices) one. Then each edge in the subgraph induced by $U$ is contained in at least two vertices of $U$, as otherwise we could remove a vertex of degree one or zero in order to obtain a 1-dense set $U' \subset U$. This implies that the core of $H_{n,cn,k}$ contains all minimal 1-dense subsets. In the remainder of this section we therefore focus on the analysis of the core of $H_{n,cn,k}$, and we will show that with probability $1 - o(1)$ it does not contain any 1-dense subset.

Proposition 3.1 and Theorem 3.2 guarantee that we can perform all calculations in the Poisson cloning model, i.e., it is sufficient to consider the core of $\widetilde{H}_{n,p,k}$, where $p = ck/\binom{n-1}{k-1}$, and $c = c_k^* - \gamma < 1$, where $\gamma > 0$ will be determined later. Let $C = C(\widetilde{H}_{n,p,k})$ denote the core of $\widetilde{H}_{n,p,k}$. For notational convenience we use the symbols $n_2$ and $m_2$ to denote the number of vertices and the number of edges, respectively, of C. Our proof then proceeds by exposing C in the following three stages.

1. We will expose the values of $n_2$ and $\Lambda_{c,k}$ and condition on the event that they obtain specific values within the ranges stated in Theorem 3.3.

2. We will expose the degrees $\mathbf{d} = (d_1, \ldots, d_{n_2})$ of the vertices of the core, which, again according to Theorem 3.3, are distributed as i.i.d. 2-truncated Poisson random variables with parameter $\Lambda_{c,k}$.

3. We expose $H_{\mathbf{d},k}$, i.e., the edges between the clones that emerge from the degree sequence that was exposed in Stage 2.

In order to obtain tight bounds for the probability that there are 1-dense subsets in the core of $\widetilde{H}_{n,p,k}$, we will exploit more sophisticated properties of such sets. In particular, we will use the following statement, which was observed by Bohman and Kim [4]. We present a proof for the sake of completeness.

**Proposition 4.3.** *Let $H$ be a $k$-graph with density $< 1$ and let $U$ be an inclusion maximal 1-dense subset of $V_H$. Then $e_U = v_U$ and all edges $e \in E_H$ satisfy $|e \cap U| \neq k - 1$.*

*Proof.* If $e_U > v_U$, then let $U' = U \cup \{v\}$, where $v$ is any vertex in $V_H \setminus U$. Note that such a vertex always exists, as $U \neq V_H$. Moreover, denote by $d$ the degree of $v$ in $U$, i.e., the number of edges in $H$ that contain $v$ and all other vertices only from $U$. Then

$$\frac{e_{U'}}{v_{U'}} = \frac{e_U + d}{v_U + 1} \geq \frac{e_U}{v_U + 1} \geq 1,$$

which contradicts the maximality of $U$. Similarly, if there was an edge $e$ such that $|e \cap U| = k - 1$, then we could construct a larger 1-dense subset of $V_H$ by adding the vertex in $e \setminus U$ to $U$. □

The following lemma bounds the probability that a given set of the core is maximal 1-dense, *assuming that the degree sequence has been exposed*. That is, the randomness is that of the 3rd stage described above. However, the calculations rely neither on the certain values of $n_2$ and $\Lambda_{c,k}$ which are determined by Theorem 3.3 nor on the distribution of the degrees which is obtained through the Poisson cloning model, but it is much more general. A similar statement was shown in [4] for the special case $k = 4$, and the proof is inspired from there.

**Lemma 4.4.** *Let $k \geq 2$, $\mathbf{d} = (d_1, \ldots, d_N)$ be a degree sequence and $U \subseteq \{1, \ldots, N\}$. Moreover, set $m = k^{-1} \sum_{i=1}^{N} d_i$ and $q = (km)^{-1} \sum_{i \in U} d_i$. Assume that $m < N$. If $\mathcal{B}_U$ denotes the event that $U$ is an inclusion maximal 1-dense set of $H_{\mathbf{d},k}$, then*

$$\mathbb{P}\left(\mathcal{B}_U\right) \leq (1 + o(1))\sqrt{2m}\binom{m}{|U|}(2^k - k - 1)^{m-|U|} \cdot e^{-kmH(q)},$$

*where $H(x) = -x \ln x - (1 - x) \ln(1 - x)$ denotes the entropy function.*

*Proof.* Recall that $H_{\mathbf{d},k}$ is obtained by beginning with $d_i$ clones for each $1 \leq i \leq N$ and by choosing uniformly at random a perfect $k$-matching on this set of clones. This is equivalent to throwing $km$ balls into $m$ bins such that every bin contains $k$ balls. In order to estimate the probability for $\mathcal{B}_U$ assume that we color the $kqm$ clones of the vertices in $U$ with red, and the remaining $k(1-q)m$ clones with blue. So, by applying Proposition 4.3 we are interested in the probability for the event that there are exactly $|U|$ bins with $k$ red balls and no bin that contains exactly one blue ball.

We estimate the above probability as follows. We begin by putting into each bin $k$ *black* balls, labeled with the numbers $1, \ldots, k$. Let $\mathcal{K} = \{1, \ldots, k\}$, and let $X_1, \ldots, X_m$ be independent random sets such that for $1 \leq i \leq m$

$$\forall \mathcal{K}' \subseteq \mathcal{K} \; : \; \mathbb{P}\left(X_i = \mathcal{K}'\right) = q^{|\mathcal{K}'|}(1-q)^{k-|\mathcal{K}'|}.$$

Note that $|X_i|$ is the binomial distribution $\text{Bin}(k, q)$. We then recolor the balls in the $i$th bin that are in $X_i$ with red, and all others with blue. So, the total number of red balls is $X = \sum_{i=1}^m |X_i|$, and it follows that the number of ways to partition $kqm$ red and $k(1-q)m$ blue balls in $m$ bins such that each bin has $k$ balls is $\mathbb{P}(X = kqm) \cdot 2^{km}$. Consequently, overall there are

$$Z = \mathbb{P}(X = kqm) \, 2^{km} \cdot (kqm)! \cdot (k(1-q)m)! \cdot (k!)^{-m}$$

ways to throw $kqm$ red and $k(1-q)m$ blue balls in $m$ bins such that each bin has $k$ balls. Note that $\mathbb{E}(X) = kqm$, and that $X$ is distributed like $\text{Bin}(km, q)$. A straightforward application of Stirling's Formula gives then

$$\mathbb{P}(X = kqm) = \mathbb{P}(X = \mathbb{E}(X)) = (1 + o(1))(2\pi q(1-q)km)^{-1/2}.$$

Let $R_j$ be the number of $X_i$'s that contain $j$ elements. Then, the probability that there are exactly $|U|$ bins with $k$ red balls and no bin that contains exactly one blue ball is precisely

$$P = \mathbb{P}(X = kqm \wedge R_k = |U| \wedge R_{k-1} = 0) \, 2^{km} \cdot (kqm)! \cdot (k(1-q)m)! \cdot (k!)^{-m}.$$

Using this notation we may estimate

$$\mathbb{P}(\mathcal{B}_U) = \frac{P}{Z} \leq (1 + o(1))\sqrt{2m} \cdot \mathbb{P}(X = kqm \wedge R_k = |U| \wedge R_{k-1} = 0). \tag{4.2}$$

Let $p_j = \mathbb{P}(|X_i| = j) = \binom{k}{j} q^j (1-q)^{k-j}$. Moreover, define the set of integer sequences

$$\mathcal{A} = \left\{ (b_0, \ldots, b_{k-2}) \in \mathbb{N}^{k-1} \; : \; \sum_{j=0}^{k-2} b_j = m - |U| \text{ and } \sum_{j=0}^{k-2} j b_j = kqm - k|U| \right\}.$$

Then

$$\mathbb{P}(X = kqm \wedge R_k = |U| \wedge R_{k-1} = 0) = \sum_{(b_0, \ldots, b_{k-2}) \in \mathcal{A}} \binom{m}{b_0, \ldots, b_{k-2}, 0, |U|} \cdot \prod_{j=0}^{k-2} p_j^{b_j} \cdot p_k^{|U|}.$$

Now observe that the summand can be rewritten as

$$\binom{m}{|U|} q^{kqm} (1-q)^{k(1-q)m} \cdot \binom{m - |U|}{b_0, \ldots, b_{k-2}} \prod_{j=0}^{k-2} \binom{k}{j}^{b_j}$$

Thus, by using (4.2) we infer that

$$\mathbb{P}(\mathcal{B}_U) \leq (1 + o(1))\sqrt{2m} \binom{m}{|U|} q^{kqm} (1-q)^{k(1-q)m} (2^k - k - 1)^{m - |U|} \cdot S,$$

where

$$S = \sum_{(b_0, \ldots, b_{k-2}) \in \mathcal{A}} \binom{m - |U|}{b_0, \ldots, b_{k-2}} \prod_{j=0}^{k-2} \left( \frac{\binom{k}{j}}{2^k - k - 1} \right)^{b_j} \leq 1.$$

□

As already mentioned, the above lemma gives us a bound on the probability that a subset of the core with a given number of vertices and total degree is maximal 1-dense, assuming that the degree sequence is given. Particularly, it exploits the randomness that is present in the 3rd stage of the exposure process defined in the beginning of Section 4.1. In order to show that the core contains no 1-dense subset, we will estimate the number of such subsets using the first two stages of the exposure strategy.

For a positive integer $t$ and a $q \in [0,1]$ let $X_{q,t} = X_{q,t}(C)$ denote the number of subsets of C with $t$ vertices and total degree $d = \lfloor q \cdot km_2 \rfloor$, where $m_2$ is the number of edges in C. (For the sake of the simplicity we will say that the number of vertices of a set is its *size* and its total degree is its *degree*. Moreover, we will omit writing "$\lfloor . \rfloor$" from now on.) Note that $X_{q,t}$ is a random variable that depends *only* on the outcomes of the first two stages of the exposure of the core. Let also $X_{q,t}^{(1)}$ denote the subset of these sets that are maximal 1-dense. Then the following corollary is an immediate consequence of Markov's inequality.

**Corollary 4.5.** *Let $U$ be an arbitrary set of vertices of C of size $t$ and degree $d = q \cdot km_2$, and let $\mathcal{B}_U$ be defined as in Lemma 4.4. Then*

$$\mathbb{P}\left(X_{q,t}^{(1)} > 0 \mid X_{q,t}\right) \leq X_{q,t}\mathbb{P}\left(\mathcal{B}_U\right).$$

Let $\delta > 0$. Working on the probability space of the first two stages of our exposure process, we will estimate the expected value of $X_{q,t}$ and thereafter, using Markov's inequality, we will show that $X_{q,t}$ itself is with probability $1 - o(1)$ no more that its expectation multiplied by a polynomial factor. Recall that $n_2$ denotes the number of vertices in the core of $\widetilde{H}_{n,p,k}$. By applying Theorem 3.3 we obtain that with probability $1 - o(1)$

$$n_2 = n(1 - e^{-\xi} - \xi e^{-\xi}) \pm \delta^2 n, \text{ and } \Lambda_{c,k} = \xi \pm \delta^2,$$

where $\xi = \bar{x}ck$ and $\bar{x}$ is the largest solution of $x = (1 - e^{-xck})^{k-1}$. As $\xi$ is increasing with respect to $c$ (cf. Claim 3.5), there exists a $\gamma = \gamma(\delta) > 0$ such that $c = c_k^* - \gamma$ and $\xi = \xi^* - \delta$, where $\xi^*$ is the unique solution of $k = \frac{\xi^*(e^{\xi^*}-1)}{e^{\xi^*}-1-\xi^*}$. Therefore

$$n_2 = n(1 - e^{-\xi^*} - \xi^* e^{-\xi^*}) \pm O(\delta n), \text{ and } \Lambda_{c,k} = \xi^* - \delta \pm \delta^2.$$

We shall be using these facts without further reference. This is all we need from the randomness contained in the first stage of our exposure process. On the probability space of the second stage, we will condition on an event $\mathcal{E}$, which informally says that the sum of degrees of the vertices does not deviate much from its expected value. More specifically, recall that the degrees of the vertices of C are i.i.d. 2-truncated Poisson random variables with parameter $\Lambda_{c,k}$. Therefore, the expected value of the sum of the degrees is equal to

$$\bar{D} := n_2 \frac{\Lambda_{c,k}(1 - e^{-\Lambda_{c,k}})}{1 - e^{-\Lambda_{c,k}} - \Lambda_{c,k}e^{-\Lambda_{c,k}}} = n_2(k(1 - e_k\delta) \pm \Theta(\delta^2)) \tag{4.3}$$

for some constant $e_k > 0$. Now, let $d_1, \ldots, d_{n_2}$ denote the random variables that are the degrees of the vertices of C, and let $\mathcal{E}$ be the event "$D = \sum_{i=1}^{n_2} d_i \in \bar{D} \pm n^{2/3}$". Moreover, recall that $m_2$ denotes the number of edges in C, i.e., $m_2 = D/k$. As we observed in the proof of Corollary 3.4, $\text{Var}(D) = \Theta(n_2)$ and, therefore, Chebyschev's inequality yields:

$$\mathbb{P}\left(\mathcal{E}\right) = 1 - O\left(n^{-1/3}\right). \tag{4.4}$$

Note that on the event $\mathcal{E}$

$$m_2 = n_2(1 - e_k\delta + \Theta(\delta^2)), \text{ where } e_k \text{ is given by } \frac{\xi(e^\xi - 1)}{e^\xi - \xi - 1} = k(1 - e_k\delta + \Theta(\delta^2)). \quad (4.5)$$

Let $\mathcal{E}_1$ be the event that conditional on $\mathcal{E}$ we have $X_{q,t} \leq n^2 \mathbb{E}(X_{q,t} \mid \mathcal{E})$. Markov's inequality immediately implies that $\mathbb{P}(\mathcal{E}_1 \mid \mathcal{E}) \geq 1 - 1/n^2$. So, Corollary 4.5 yields the following.

**Corollary 4.6.** *Let $U$ be an arbitrary set of vertices of $C$ of size $t$ and degree $d = q \cdot km_2$, and let $\mathcal{B}_U$ be defined as in Lemma 4.4. Then*

$$\mathbb{P}\left(X_{q,t}^{(1)} > 0\right) \leq n^2 \mathbb{E}(X_{q,t} \mid \mathcal{E})\mathbb{P}(\mathcal{B}_U) + o(1).$$

*Proof.* We have

$$\mathbb{P}\left(X_{q,t}^{(1)} > 0\right) \leq \mathbb{P}\left(X_{q,t}^{(1)} > 0 \mid \mathcal{E}_1 \cap \mathcal{E}\right) + \mathbb{P}\left(\overline{\mathcal{E}_1}\right) + \mathbb{P}\left(\overline{\mathcal{E}}\right) = \mathbb{P}\left(X_{q,t}^{(1)} > 0 \mid \mathcal{E}_1 \cap \mathcal{E}\right) + o(1).$$

Note that any event about $X_{q,t}^{(1)}$ given the value of $X_{q,t}$ is independent of $\mathcal{E}$, as this regards only the randomness in the 3rd stage of the exposure process. The proof finishes by applying Corollary 4.5. □

Let us proceed with the estimation of $\mathbb{E}(X_{q,t} \mid \mathcal{E})$. We will use the following theorem, which is a special case of Theorem II.4.I in [13].

**Theorem 4.7.** *Let $X$ be a random variable taking real values and set $c(t) = \ln \mathbb{E}(e^{tX})$, for any $t > 0$. For any $z > 0$ we define $I(z) = \sup_{t \in \mathbb{R}}\{zt - c(t)\}$. If $X_1, \ldots, X_s$ are i.i.d. random variables distributed as $X$, then for $s \to \infty$*

$$\mathbb{P}\left(\frac{\sum_{i=1}^s X_i}{s} \leq z\right) \leq \exp\left(-s \inf\{I(x) : x \leq z\}(1 + o(1))\right).$$

*The function $I(z)$ is non-negative and convex.*

**Remark** By the large deviation theory the above bound is asymptotically (i.e., for large $s$) tight. See [13] or [7] for a more detailed treatment.

With this theorem at hand we are ready to present our main tool for estimating $\mathbb{E}(X_{q,t} \mid \mathcal{E})$.

**Lemma 4.8.** *Let $T_z$ be the unique solution of $z = \frac{T_z(1-e^{-T_z})}{1-e^{-T_z}-T_z e^{-T_z}}$, where $z > 2$. Moreover, let*

$$I(z) = z(\ln T_z - \ln \xi) - \ln\left(e^{T_z} - T_z - 1\right) + \ln\left(e^\xi - \xi - 1\right), \quad (4.6)$$

*and set $I(2) := \ln 2 - 2\ln\xi + \ln(e^\xi - \xi - 1)$. Then $I(z)$ is continuous for all $z \geq 2$ and convex. It has a unique minimum at $\mu = \frac{\xi(e^\xi-1)}{e^\xi-\xi-1} = k(1 - e_k\delta + \Theta(\delta^2))$, where $I(\mu) = 0$.*

*Let $X_1, \ldots, X_s$ be i.i.d. 2-truncated Poisson random variables with parameter $\Lambda_{c,k}$. Then, uniformly for any $z$ such that $2 \leq z \leq \mu$, we have*

$$\mathbb{P}\left(\frac{\sum_{i=1}^s X_i}{s} < z\right) \leq \exp\left(-sI(z)(1 + o(1)) + O(s\delta^2)\right),$$

*as $s \to \infty$.*

*Proof.* We shall first calculate $c(t) = \ln \mathbb{E}(e^{tX})$, where $X$ is a random variable which follows a 2-truncated Poisson distribution with parameter $\xi$. Note that

$$\exp\{c(t)\} = \mathbb{E}(e^{tX}) = \frac{1}{1 - e^{-\xi} - \xi e^{-\xi}} \sum_{\ell=2}^{\infty} e^{-\xi} \frac{e^{t\ell} \xi^\ell}{\ell!} = \frac{e^{\xi e^t} - \xi e^t - 1}{e^\xi - \xi - 1}.$$

Differentiating $tz - c(t)$ with respect to $t$ we obtain

$$(tz - c(t))' = \left(tz - \ln\left(\frac{e^{\xi e^t} - \xi e^t - 1}{e^\xi - \xi - 1}\right)\right)'$$

$$= z - \frac{e^\xi - \xi - 1}{e^{\xi e^t} - \xi e^t - 1} \frac{\left(\xi e^t \left(e^{\xi e^t} - 1\right)\right)}{e^\xi - \xi - 1} = z - \frac{\xi e^t \left(e^{\xi e^t} - 1\right)}{e^{\xi e^t} - \xi e^t - 1}.$$

Setting $T := \xi e^t$ yields the compact form

$$(tz - c(t))' = z - \frac{T\left(e^T - 1\right)}{e^T - T - 1}.$$

As this suggests, it will be convenient to parameterize $zt - c(t)$ in terms of $T$.

Setting the derivate to 0, we obtain a unique $T$ that solves the above and which we denote $T_z$. The uniqueness of the solution for $z > 2$ follows from the fact that the function $x(e^x - 1)/(e^x - x - 1)$ is strictly increasing with respect to $x$ and, as $x$ approaches 0, it tends to 2. In other words, $T_z$ is the unique positive real number that satisfies

$$z = \frac{T_z \left(e^{T_z} - 1\right)}{e^{T_z} - T_z - 1}. \tag{4.7}$$

Letting $t_z$ be such that $T_z = \xi e^{t_z}$, we obtain:

$$-c(t_z) = -\ln(e^{T_z} - T_z - 1) + \ln(e^\xi - \xi - 1)$$

and

$$t_z z = z(\ln T_z - \ln \xi).$$

The function $-c(t)$ is concave with respect to $t$ (cf. Proposition VII.1.1 in [13, p. 229]); also adding the linear term $zt$ does preserve concavity. So $t_z$ is the point where the unique maximum of $zt - c(t)$ is attained over $t \in \mathbb{R}$. Therefore,

$$I(z) = z(\ln T_z - \ln \xi) - \ln(e^{T_z} - T_z - 1) + \ln(e^\xi - \xi - 1).$$

For $z = \frac{\xi(e^\xi - 1)}{e^\xi - 1 - \xi}$, we have $T_z = \xi$ and therefore the above equality yields $I(\mu) = 0$.

As far as $I(2)$ is concerned, note that strictly speaking this is not defined, as there is no positive solution of the equation $2 = T(e^T - 1)/(e^T - T - 1)$. However, the function on the right-hand side converges to 2 as $T \to 0$ from the right. So, we can express $I(2)$ as a limit:

$$I(2) := \lim_{T \to 0^+} \left(2 \ln T - \ln(e^T - T - 1)\right) - 2 \ln \xi + \ln(e^\xi - \xi - 1).$$

But

$$\lim_{T\to 0^+} \left(2\ln T - \ln(e^T - T - 1)\right) = \lim_{T\to 0^+} \ln \frac{T^2}{e^T - T - 1} = \lim_{T\to 0^+} \ln \frac{T^2}{\frac{T^2}{2} + \frac{T^3}{3!} + \cdots}$$

$$= \lim_{T\to 0^+} \ln \frac{1}{\frac{1}{2} + \frac{T}{3!} + \cdots} = \ln 2,$$

and therefore

$$I(2) = \ln 2 - 2\ln \xi + \ln(e^\xi - \xi - 1).$$

All the above analysis was for a 2-truncated Poisson variable $X$ with parameter $\xi$. In order to obtain the statement of the lemma, recall that $\Lambda_{c,k} = \xi \pm \delta^2$, and that the $X_i$'s are 2-truncated Poisson variables with parameter $\Lambda_{c,k}$. A straightforward argument by using Taylor's theorem then shows that $\tilde{c}(t) := \ln \mathbb{E}(e^{tX_1}) = c(t) + O(\delta^2)$, which holds uniformly for $t$ in any bounded domain. Moreover, set $\tilde{I}(z) := \sup_{t\in\mathbb{R}}\{zt - \tilde{c}(t)\}$, for all $z > 2$. Therefore uniformly for all $z > 2$, $\tilde{I}(z) = I(z) + O(\delta^2)$.

Also, according to Theorem 4.7 the function $I(z)$ is non-negative and convex on its domain. So if $z \leq \mu$, then $\inf\{I(x) : x \leq z\} = I(z)$ and the second part of the lemma follows. $\square$

The following lemma bounds the expected value of $X_{q,t}$.

**Lemma 4.9.** *Let $t = \beta n_2$ and $q \geq \beta$. Then, for any $n_2$ sufficiently large and $\delta > 0$ sufficiently small*

$$\mathbb{E}(X_{q,t} \mid \mathcal{E}) \leq 2\binom{n_2}{t} \exp\left(-n_2(1-\beta)I\left(\frac{k(1-q)}{1-\beta}\right)(1+o(1)) + O(n_2\delta^2)\right),$$

*where $I(z)$ is given in (4.6).*

*Proof.* There are $\binom{n_2}{t}$ ways to select a set with $t$ vertices. We shall next calculate the probability that one of them has the claimed property, and the statement will follow from a simple union bound. Let $U$ be a fixed subset of the vertex set of C that has size $t$ and let $d_1, \ldots, d_t$ denote the random variables that are degrees of the vertices in $U$. Thus, we want to estimate the probability of the event $\sum_{i=1}^{t} d_i = q \cdot km_2$ conditional on $\mathcal{E}$. Let $d_{t+1}, \ldots, d_{n_2}$ denote the random variables that are the degrees of the vertices which do not belong to $U$; these are i.i.d. 2-truncated Poisson variables with parameter $\Lambda_{c,k} = \xi \pm \delta^2$. If $\mathcal{E}$ is realized, then recall (4.5) and note that

$$\frac{\sum_{i=t+1}^{n_2} d_i}{n_2 - t} = \frac{km_2 - qkm_2}{n_2 - t} = \frac{k(1-q)m_2}{(1-\beta)n_2} \leq \frac{k(1-q)}{1-\beta}(1 - e_k\delta + \Theta(\delta^2)).$$

Note that the last expression is at most $k(1 - e_k\delta + \Theta(\delta^2)) = \frac{\xi(e^\xi - 1)}{e^\xi - \xi - 1} + \Theta(\delta^2) = \mu + \Theta(\delta^2)$. So, by applying Lemma 4.8, using the convexity of $I(z)$, and the fact $\mathbb{P}(\mathcal{E}) = 1 - o(1)$ the proof is completed. $\square$

Corollary 4.6 along with Lemmas 4.4 and 4.9 yield the following estimate.

**Lemma 4.10.** *Let $t = \beta n_2$ and $q \geq \beta$, and let $m_2 = (1 - e_k\delta \pm \Theta(\delta^2))n_2$. Then*

$$\mathbb{P}\left(X_{q,t}^{(1)} > 0\right) = o(1) +$$

$$\binom{n_2}{t}\binom{m_2}{t}(2^k - k - 1)^{m_2 - t} \exp\left(-km_2 H(q) - n_2(1-\beta)I\left(\frac{k(1-q)}{1-\beta}\right) + O(n_2\delta^2)\right).$$

*Moreover, when $q < \beta$, the above probability is 0.*

The rest of this section is devoted to the proof of the following statement.

**Lemma 4.11.** *Let $m_2 = (1 - e_k\delta \pm \Theta(\delta^2))n_2$, where $\delta > 0$ is sufficiently small. Then the following holds with probability $1 - o(1)$. For any $t = \beta n_2$ where $0.7 \leq \beta \leq 1 - e_k\delta/2$ and any $q$ such that $\beta \leq q \leq 1 - \frac{2(1-\beta)}{k}$ we have $X_{q,t}^{(1)} = 0$.*

The lower bound for $q$ in the above lemma comes from the fact that otherwise the corresponding probability is zero, cf. Lemma 4.10. Moreover, the upper bound in the range of $q$ stems from the fact that the average degree of the complement of a set with $t$ vertices and total degree $q \cdot km_2$ is at least two. More precisely, the total degree of the core satisfies for small $\delta > 0$

$$km_2 \geq q \cdot km_2 + 2(n_2 - t) \implies q \leq 1 - \frac{2(1-\beta)n_2}{km_2} \stackrel{(4.5)}{\leq} 1 - \frac{2(1-\beta)}{k}.$$

With the above result at hand we can finally complete the proof of Theorem 2.3.

*Proof of Theorem 2.3.* Recall that it is sufficient to show that the core C of $\widetilde{H}_{n,p,k}$ contains with probability $1 - o(1)$ no maximal 1-dense subsets (see also the discussion in the beginning of Section 4.1), where $p = ck/\binom{n-1}{k-1}$. Note also that it is sufficient to argue about subsets of size up to, say, $(1 - e_k\delta/2)n_2$, where $n_2$ is the number of vertices in C, as (4.5) implies for small $\delta$ that all larger subsets have density smaller than 1.

Let $k \geq 5$. By applying Lemma 4.1 we obtain that $H_{n,cn,k}$ does not obtain any 1-dense set with less that $0.7n$ vertices, and the same is true for $\widetilde{H}_{n,p,k}$, by Proposition 3.1 and Theorem 3.2. In particular, C does not contain such a subset, and it remains to show the claim for sets of size at least $0.7n \geq 0.7n_2$. The proof is completed by applying Lemma 4.11, as we can choose $\delta > 0$ as small as we please.

The case $k = 3$ requires slightly more work. Lemma 4.2 guarantees that C has no subset with $\leq n/2$ vertices that contains exactly as many edges as vertices, and there is no edge that contains precisely two vertices in that set. In other words, by using Proposition 4.3, C does not contain a maximal 1-dense set with $n/2$ vertices. However, we know that

$$n_2 = (1 - e^{\xi^*} - \xi^* e^{-\xi^*} \pm O(\delta))n, \text{ where } 3 = \frac{\xi^*(e^{\xi^*} - 1)}{e^{\xi^*} - 1 - \xi^*}.$$

Numerical calculations imply that $n_2 \geq 0.63n$ for any $\delta$ that is small enough. So, C does not contain any maximal 1-dense subset with less than $n/2 \leq n_2/(2 \cdot 0.63) \leq 0.77n_2$ vertices. In this case, the proof is completed again by applying Lemma 4.11.

Finally, the case $k = 4$ was treated in [4]. This completes the proof of Theorem 2.3. □

## 4.2 Proof of Lemma 4.11

Let
$$f(\beta, q) := 2\, H(\beta) + (1 - \beta)\ln(2^k - k - 1) - kH(q) - (1 - \beta)I\left(\frac{k(1-q)}{1-\beta}\right).$$

By using Lemma 4.10 and the well-known bounds $\binom{n_2}{t} \leq e^{n_2 H(\beta)}$ and $\binom{m_2}{t} \leq \binom{n_2}{t}$ we infer that
$$\frac{1}{n_2}\ln \mathbb{P}\left(X_{q,t}^{(1)} > 0\right) \leq f(\beta, z) + e_k\delta\left(kH(q) - \ln(2^k - k - 1)\right) + O(\delta^2).$$

We will show the following.

**Claim 4.12.** *There exists a $C > 0$ such that for any small enough $\varepsilon > 0$ the following is true. Let $0.7 \leq \beta \leq 1 - \varepsilon$, and $q$ as in Lemma 4.11. Then*

$$f(\beta, q) \leq -C\varepsilon + O(\delta^2).$$

This completes the proof of Lemma 4.11 as follows. We distinguish between the following cases. First, note that if $0.7 \leq \beta \leq 1 - \sqrt{\delta}$, then the above claim yields for sufficiently small $\delta > 0$

$$\frac{1}{n_2} \ln \mathbb{P}\left(X_{q,t}^{(1)} > 0\right) \leq -C\sqrt{\delta} + O(\delta) \leq -C\sqrt{\delta}/2.$$

This implies that with probability $1 - e^{-\Omega(\sqrt{\delta}n_2)}$ we have $X_{q,t}^{(1)} = 0$ for all $t = \beta n_2$ and $q$.

Finally, if $1 - \sqrt{\delta} \leq \beta \leq 1$, then the above claim implies that $f(\beta, q) = O(\delta^2)$. Moreover, by the monotonicity of the entropy function and $q \geq \beta$ we have for sufficiently small $\delta > 0$

$$kH(q) - \ln(2^k - k - 1) \leq kH(0.99) - \ln(2^k - k - 1).$$

A simple calculation and the fact $H(0.99) < 0.06$ show that the above expression is negative for all $k \geq 3$. This completes the proof also in this case.

The rest of the paper is devoted to the proof of Claim 4.12. We proceed as follows. We will fix arbitrarily a $\beta$ and we will consider $f(\beta, z)$ solely as a function of $z$. Then we will show that if $q_0 = q_0(\beta)$ is a point inside the domain where $\partial f/\partial q = 0$, then $f(\beta, q_0) \leq -C_1\varepsilon$. Additionally, we will show that this holds for $f(\beta, \beta)$ and $f\left(\beta, 1 - \frac{2(1-\beta)}{k}\right)$.

**Bounding $f(\beta, q)$ at its critical points**

Let $\beta$ be fixed. We will evaluate $f(\beta, q)$ at a point where the partial derivative with respect to $q$ vanishes. To calculate the partial derivative with respect to $q$, we first need to determine the derivative of $I(z)$ with respect to $z$.

According to Lemma 4.8, $I(z) = z(\ln T_z - \ln \xi) - \ln(e^{T_z} - T_z - 1) + \ln(e^{\xi} - \xi - 1)$. So differentiating this with respect to $z$ we obtain:

$$I'(z) = \ln T_z - \ln \xi + z\frac{1}{T_z}\frac{dT_z}{dz} - \frac{e^{T_z} - 1}{e^{T_z} - T_z - 1}\frac{dT_z}{dz} \stackrel{(4.7)}{=} \ln T_z - \ln \xi. \tag{4.8}$$

However, in the differentiation of $f$ we need to differentiate $I(k(1-q)/(1-\beta))$ with respect to $q$. Using (4.8), we obtain

$$\frac{\partial I\left(\frac{k(1-q)}{1-\beta}\right)}{\partial q} = -\frac{k}{1-\beta}(\ln H_q - \ln \xi),$$

where $H_q$ is the unique solution of the equation

$$\frac{k(1-q)}{1-\beta} = \frac{H_q(e^{H_q} - 1)}{e^{H_q} - H_q - 1}.$$

(Observe that the choice of the range of $q$ is such that the left-hand side of the above equation is at least 2. So, $H_q$ is well-defined.) Also, an elementary calculation shows that $H'(q) = \ln\left(\frac{1-q}{q}\right)$. All the above facts together yield the derivative of $f(\beta, q)$ with respect to $q$:

$$\frac{\partial f(\beta, q)}{\partial q} = k\left(-\ln\left(\frac{1-q}{q}\right) + (\ln H_q - \ln \xi)\right).$$

Therefore, if $q_0$ is a critical point, that is, if $\left.\frac{\partial f(\beta,q)}{\partial q}\right|_{q=q_0} = 0$, then $q_0$ satisfies

$$T_0 = \xi \frac{1-q_0}{q_0}, \quad \text{where} \quad \frac{k(1-q_0)}{1-\beta} = \frac{T_0(e^{T_0}-1)}{e^{T_0} - T_0 - 1}. \tag{4.9}$$

At this point, we have the main tool that will allow us to evaluate $f(\beta, q_0)$. We will use (4.9) in order to eliminate $T_0$ and express $f(\beta, q_0)$ solely as a function of $q_0$.

**Claim 4.13.** *For any given $\beta$, if $q_0 = q_0(\beta)$ is a critical point of $f(\beta, q)$ with respect to $q$, then*

$$\begin{aligned}f(\beta,q_0) = {}& 2H(\beta) + (1-\beta)\ln\left(\frac{2^k - k - 1}{e^\xi - \xi - 1}\right) + k\ln(q_0) \\ & + (1-\beta)\left(2\ln\xi + \ln(1-q_0) - \ln(q_0) + \ln(1-\beta) - \ln(kq_0 - \xi(1-\beta))\right).\end{aligned} \tag{4.10}$$

*Proof of Claim.* Firstly note that

$$\begin{aligned}I\left(\frac{k(1-q_0)\beta}{1-\beta}\right) &= \frac{k(1-q_0)\beta}{1-\beta}\ln\frac{T_0}{\xi} - \ln\left(e^{T_0} - T_0 - 1\right) + \ln\left(e^\xi - \xi - 1\right)\\ &\stackrel{(4.9)}{=} \frac{k(1-q)\beta}{1-\beta}\ln\left(\frac{1-q_0}{q_0}\right) - \ln\left(e^{T_0} - T_0 - 1\right) + \ln\left(e^\xi - \xi - 1\right),\end{aligned}$$

whence

$$\begin{aligned}-(1-\beta)I\left(\frac{k(1-q_0)\beta}{1-\beta}\right) &= -k(1-q_0)\ln\left(\frac{1-q_0}{q_0}\right) + (1-\beta)\ln\left(\frac{e^{T_0} - T_0 - 1}{e^\xi - \xi - 1}\right)\\ &= -k(1-q_0)\ln(1-q_0) + k\ln(q_0) - kq_0\ln(q_0) + (1-\beta)\ln\left(\frac{e^{T_0} - T_0 - 1}{e^\xi - \xi - 1}\right).\end{aligned}$$

Also, the definition of the entropy function implies that

$$-kH(q_0) = kq_0\ln(q_0) + k(1-q_0)\ln(1-q_0).$$

Thus

$$-(1-\beta)I\left(\frac{k(1-q_0)\beta}{1-\beta}\right) - kH(q_0) = k\ln(q_0) + (1-\beta)\ln\left(\frac{e^{T_0} - T_0 - 1}{e^\xi - \xi - 1}\right). \tag{4.11}$$

Now we will express $e^{T_0}$ as a rational function of $T_0$ and $z_0$. Setting $\hat{z} := \frac{k(1-q_0)}{1-\beta}$ and solving (4.9) with respect to $e^{T_0}$ yields

$$e^{T_0} = \frac{\hat{z} + T_0(\hat{z} - 1)}{\hat{z} - T_0}.$$

Therefore

$$e^{T_0} - 1 - T_0 = \frac{\hat{z} + T_0\hat{z} - T_0 - \hat{z} + T_0 - T_0\hat{z} + T_0^2}{\hat{z} - T_0} = \frac{T_0^2}{\hat{z} - T_0}.$$

Note that

$$\hat{z} - T_0 = \frac{k(1-q_0)}{1-\beta} - \frac{\xi(1-q_0)}{q_0} = \frac{(1-q_0)(kq_0 - \xi(1-\beta))}{(1-\beta)q_0}.$$

Thus we obtain:

$$\ln(e^{T_0} - 1 - T_0) = 2\ln T_0 - \ln(\hat{z} - T_0)$$
$$\stackrel{(4.9)}{=} 2\ln\xi + 2\ln(1-q_0) - 2\ln(q_0) - \ln(1-q_0) - \ln(kq_0 - \xi(1-\beta))$$
$$+ \ln q_0 + \ln((1-\beta))$$
$$= 2\ln\xi + \ln(1-q_0) - \ln(q_0) + \ln(1-\beta) - \ln(kq_0 - \xi(1-\beta)).$$

Substituting this into (4.11) and adding the remaining terms, we obtain (4.10). □

We will now treat $q_0$ as a free variable lying in the interval where $q$ lies into, and we will study $f(\beta, q_0)$ for a fixed $\beta$ as a function of $q_0$. In particular, we will show that for any fixed $\beta$ in the domain of interest $f(\beta, q_0)$ is increasing. Thereafter, we will evaluate $f(\beta, q_0)$ at the largest possible value that $q_0$ can take, which is $1 - \frac{2(1-\beta)}{k}$, and show that this value is negative.

**Claim 4.14.** *For any $k \geq 3$ and for any $\beta \geq 0.7$ we have*

$$\frac{\partial f(\beta, q_0)}{\partial q_0} > 0.$$

*Proof of Claim.* The partial derivative of $f(\beta, q_0)$ with respect to $q_0$ is

$$\frac{\partial f(\beta, q_0)}{\partial q_0} = \frac{k}{q_0} - \frac{1-\beta}{1-q_0} - \frac{1-\beta}{q_0} - \frac{k(1-\beta)}{kq_0 - \xi(1-\beta)}.$$

Since $q_0 \leq 1 - \frac{2(1-\beta)}{k}$, we obtain

$$1 - q_0 \geq \frac{2(1-\beta)}{k} \Rightarrow -\frac{1-\beta}{1-q_0} \geq -\frac{k}{2}.$$

Also $q_0 \geq \beta$ and $\xi \leq k$. Therefore,

$$kq_0 - \xi(1-\beta) \geq k\beta - k(1-\beta) = 2\beta k - k = k(2\beta - 1).$$

Substituting these bounds into the formula of $\frac{\partial f(\beta, q_0)}{\partial q_0}$ yields:

$$\frac{\partial f(\beta, q_0)}{\partial q_0} \geq \frac{k}{q_0} - \frac{k}{2} - \frac{1-\beta}{q_0} - \frac{1-\beta}{2\beta - 1}$$
$$\geq k\frac{k-1+\beta}{k-2+2\beta} - \frac{k}{2} - \frac{1-\beta}{2\beta - 1} \geq k\left(1 - \frac{1}{2} - \frac{1-\beta}{k(2\beta - 1)}\right).$$

But

$$\frac{1}{2} > \frac{1-\beta}{k(2\beta - 1)},$$

as $k(2\beta - 1) > 2(1-\beta)$, which is equivalent to $2\beta(k+1) > k+2$, or, $2\beta > \frac{k+2}{k+1} = 1 + \frac{1}{k+1}$. Since $\beta \geq 0.7$ and $k \geq 3$, the above holds. □

Now we proceed with setting $q_0 := 1 - \frac{2(1-\beta)}{k}$ into $f(\beta, q_0)$ and obtain a function which depends only on $\beta$, namely:

$$h(\beta) := 2H(\beta) + (1-\beta)\ln\left(\frac{2^k - k - 1}{e^\xi - \xi - 1}\right) + k\ln\left(\frac{k - 2 + 2\beta}{k}\right)$$
$$+ (1-\beta)\left(2\ln\xi + \ln\left(2(1-\beta)^2\right) - \ln(k - 2 + 2\beta) - \ln(k - 2 + 2\beta - \xi(1-\beta))\right).$$

We will show that $h(\beta)$ is a convex function with respect to $\beta$. Thus to obtain an upper bound on $h(\beta)$ in the interval $[0.7, 1-\varepsilon]$, it will be sufficient to evaluate $h(0.7)$ and $h(1-\varepsilon)$.

**Claim 4.15.** *For all $k \geq 3$ and all $0.7 \leq \beta \leq 1$*

$$\frac{d^2 h(\beta)}{d\beta^2} > 0.$$

*Proof.* The second derivative of $h(\beta)$ is given by

$$\frac{d^2 h(\beta)}{d\beta^2} = -\frac{2}{\beta} - \frac{4(1-\beta)}{(k+2\beta-2)^2} + \frac{2+\xi}{k-2+2\beta-\xi+\xi\beta} + \frac{(2+\xi)k}{(k-2+2\beta-\xi+\xi\beta)^2}.$$

The bounds on $\beta$ imply that

$$k - 2 + 2\beta - \xi(1-\beta) \overset{(\xi > k-1)}{<} k - 2 + 2\beta - (k-1)(1-\beta) < k\beta.$$

By substituting this bound into the expression for $d^2 h(\beta)/d\beta^2$ we obtain that

$$\frac{d^2 h(\beta)}{d\beta^2} \geq -\frac{2}{\beta} - \frac{4(1-\beta)}{(k+2\beta-2)^2} + \frac{2+\xi}{k\beta} + \frac{(2+\xi)k}{(k\beta)^2}.$$

Elementary algebra yields that

$$-\frac{2}{\beta} + \frac{2+\xi}{k\beta} + \frac{2+\xi}{k\beta^2} \overset{(\beta \leq 1)}{\geq} \frac{1}{\beta}\left(-2 + 2\frac{2+\xi}{k}\right) \overset{(\xi > k-1)}{>} \frac{1}{\beta}\left(-2 + 2\frac{k+1}{k}\right) \geq \frac{2}{k}.$$

Also for all $\beta \geq 0.7$ we have
$$\frac{4(1-\beta)}{(k+2\beta-2)^2} \leq \frac{2}{(k-1)^2}.$$

Therefore $\frac{d^2 h(\beta)}{d\beta^2} \geq \frac{2}{k} - \frac{2}{(k-1)^2} > 0$, for all $k \geq 3$. □

The above findings are summarized in the following corollary.

**Corollary 4.16.** *Let $0.7 \leq \beta \leq 1 - \varepsilon$. If $q_0$ is a critical point of $f(\beta, q)$ in the interval $k \leq q \leq 1 - \frac{2(1-\beta)}{k}$, then*
$$f(\beta, q_0) \leq \max\{h(0.7), h(1-\varepsilon)\}.$$

| $k$ | 3 | 4 | 5 | 6 |
|---|---|---|---|---|
| $h(0.7)$ | -0.02 | -0.11 | -0.2 | -0.3 |

Table 1: Upper bounds on the values of $h(\beta)$, where $\beta = 0.7$.

**Bounding $f(\beta, q)$ globally**

To conclude the proof of the lemma it suffices due to above arguments to show that for some $C > 0$
$$f(\beta, \beta), f(\beta, 1 - 2(1-\beta)/k), h(0.7), h(1-\varepsilon) \leq -C\varepsilon + O(\delta^2)$$
for all $0.7 \leq \beta \leq 1 - \varepsilon$. We begin with $h(0.7)$ and $h(1-\varepsilon)$.

**Claim 4.17.** *For all $k \geq 3$ we have $h(0.7) < -0.02$.*

*Proof.* Firstly, we give explicitly the values of $h(0.7)$ for $3 \leq k \leq 6$, see Table 1. For $k \geq 7$ we will provide an appropriate upper bound for $h(0.7)$. Note that $2 \cdot H(0.7) < 1.23$ and $2 \cdot (1 - 0.7) = 0.6$. Moreover, observe that

$$\frac{2^k - k - 1}{e^\xi - \xi - 1} \overset{\xi \geq k-1}{\leq} \frac{2^k - k - 1}{e^{k-1} - k}. \tag{4.12}$$

Additionally, note that $k \ln\left(\frac{k-2-2\beta}{k}\right) = k \ln\left(\frac{k-0.6}{k}\right) \leq -0.6$, and $(1-\beta)\ln(2(1-\beta)^2) = 0.3\ln(0.18) \leq -0.51$. So, $\xi \leq k$ implies that

$$h(0.7) \leq 0.12 + 0.3\ln\left(\frac{2^k - k - 1}{e^\xi - \xi - 1}\right) + 0.6\ln\xi - 0.3\ln\left((k - 0.6)(k - 0.6 - 0.3k)\right)$$

$$\overset{(4.12)}{\leq} 0.12 + 0.3\ln\left(\frac{2^k - k - 1}{e^{k-1} - k}\right) + 0.6\ln k - 0.3\ln\left((k - 0.6)(k - 1)0.6\right)$$

$$\leq 0.12 + 0.3\ln\left(\frac{2^k - k - 1}{e^{k-1} - k}\right) + 0.6\ln k - 0.3\ln\left(0.5k^2\right)$$

$$= 0.12 + 0.3\ln\left(\frac{2^k - k - 1}{e^{k-1} - k}\right) - 0.3\ln(1/2) \leq 0.33 + 0.3\ln\left(\frac{2^k - k - 1}{e^{k-1} - k}\right).$$

The last expression is negative for $k = 7$, and it is easily verified that it is decreasing with respect to $k$. This completes the proof. □

**Claim 4.18.** *There exists $C_1 > 0$ such that $h(1-\varepsilon) \leq -C_1\varepsilon$.*

*Proof.* Note that $h(1) = 0$ and that the left derivative of $h(\beta)$ at $\beta = 1$ does exist. The convexity of $h(\beta)$ and the fact that $h(0.7) < 0$ imply that the (left) derivative of $h(\beta)$ at $\beta = 1$ is positive. Therefore applying Taylor's Theorem on the left of $\beta = 1$ yields the claim. □

**Claim 4.19.** *For any $k \geq 3$ there is a $C_2 > 0$ such that for any $0.7 \leq \beta \leq 1 - \varepsilon$ we have $f(\beta, \beta) < -C_2\varepsilon + O(\delta^2)$.*

*Proof.* By Lemma 4.8 we have $I(\mu) = I'(\mu) = 0$ and then we observe that

$$I\left(\frac{k(1-\beta)}{1-\beta}\right) = I(k) = I(\mu(1 + O(\delta))) = I(\mu) + I'(\mu)O(\delta) + I''(\mu)O(\delta^2) = O(\delta^2).$$

So,

$$f(\beta,\beta) = -(k-2)H(\beta) + (1-\beta)\ln\left(2^k - k - 1\right) + O(\delta^2).$$

Note that for any $k \geq 3$ this function is convex with respect to $\beta$, as $-H(\beta)$ is convex and the linear term that is added does not affect its convexity. Moreover, for $\beta = 1$, we have $f(1,1) = 0$. Since $H(0.7) > 0.6$, we have

$$f(0.7, 0.7) < -(k-2) \cdot 0.6 + 0.3 \ln\left(2^k - k - 1\right).$$

The derivative of this function with respect to $k$ is $-0.6 + 0.3 \frac{2^k \ln 2 - 1}{2^k - k - 1}$. A simple calculation shows that the second summand is less than $0.35$ for all $k \geq 3$, implying that $f(0.7, 0.7)$ is decreasing with respect to $k$. So, we may set $k = 3$, thus obtaining $f(0.7, 0.7) < -0.6 + 0.3 \ln 4 < -0.1$. The above analysis along with the convexity of $f(\beta, \beta)$ finally imply with Taylor's Theorem the claimed statement. □

**Claim 4.20.** *For all $k \geq 3$ there is a $C_3 > 0$ such that for all $\beta \leq 1 - \varepsilon$*

$$f(\beta, 1 - 2(1-\beta)/k) \leq -C_3 \varepsilon.$$

*Proof.* Substituting $1 - 2(1-\beta)/k$ for $q$ into the formula of $f$ we obtain:

$$f\left(\beta, 1 - \frac{2(1-\beta)}{k}\right) = 2H(\beta) + (1-\beta)\ln(2^k - k - 1) - kH\left(\frac{k-2+2\beta}{k}\right) - (1-\beta)I(2).$$

Note that for $\beta = 1$ the expression is equal to 0. To deduce the bound we are aiming to, we will show that in fact $f(\beta, 1 - 2(1-\beta)/k)$ is an increasing function with respect to $\beta$. That is, we will show that its first derivative with respect to $\beta$ is positive. Note that by Taylor's Theorem, this implies the claim.

We get

$$\frac{\partial f\left(\beta, 1 - \frac{2(1-\beta)}{k}\right)}{\partial \beta} = 2\ln\left(\frac{1-\beta}{\beta}\right) - \ln(2^k - k - 1) - 2\ln\left(\frac{2-2\beta}{k-2+2\beta}\right) + I(2).$$

Substituting for $I(2)$ the value given in Lemma 4.8 and we obtain after some simplifications

$$\frac{\partial f\left(\beta, 1 - \frac{2(1-\beta)}{k}\right)}{\partial \beta} = \ln\left(\frac{(k - 2(1-\beta))^2(e^\xi - \xi - 1)}{2\beta^2(2^k - k - 1)\xi^2}\right).$$

We will show that the fraction inside the logarithm is greater than 1. Note first that the function $(k - 2(1-\beta))/\beta$ is decreasing with respect to $\beta$ – so we obtain a lower bound by setting $\beta = 1$. Moreover, the function $(e^x - x - 1)/x^2 = \frac{1}{2} + \frac{x}{3!} + \frac{x^2}{4!} + \cdots$ is increasing. So, as $\xi > k - 1$ we infer that $\frac{e^\xi - \xi - 1}{\xi^2} > \frac{e^{k-1} - k}{(k-1)^2}$. All these bounds together yield that

$$\frac{(k - 2(1-\beta))^2(e^\xi - \xi - 1)}{2\beta^2(2^k - k - 1)\xi^2} > \frac{1}{2}\left(\frac{k}{k-1}\right)^2 \frac{e^{k-1} - k}{2^k - k - 1}. \tag{4.13}$$

For $k = 3$ the right-hand side of the above inequality is $(1/2) \cdot (3/2)^2 \cdot ((e^2 - 3)/4) > 1$. Similarly, for $k = 4$ we obtain that it is $(1/2) \cdot (4/3)^2 \cdot ((e^3 - 4)/11) > 1$ and an analogous calculation says that it also greater than one for $k = 5$. For $k \geq 6$ note that

$$\frac{e^{k-1} - k}{2^k - k - 1} \geq \frac{e^{k-1} - k}{2^k - k} \geq \frac{1}{e} \left(\frac{e}{2}\right)^k.$$

We may omit in (4.13) the factor $k/(k-1)$ as it is greater than 1, and the right-hand side is now bounded below by $\frac{1}{2e} \left(\frac{e}{2}\right)^k$. But $(e/2)^6 > 2e$; this establishes the fact that the derivative of $f(\beta, 1 - 2(1-\beta)/k)$ with respect to $\beta$ is positive, for all $k \geq 3$. □